\documentclass[aps, prb, reprint, superscriptaddress, notitlepage, letterpaper,  10pt, floatfix, showpacs, longbibliography, balancelastpage, nofootinbib]{revtex4-2}
\pdfoutput = 1
\usepackage{amssymb, graphicx, makecell, color, xcolor, amsmath, bm, url, float, mathrsfs, braket, bbold, psfrag, dcolumn, enumitem, array, tikz, algpseudocode, setspace, adjustbox, tabularx, ragged2e, booktabs, silence, flushend, mdframed, stackengine}
\setstackEOL{\\}
\usepackage{lmodern}
\usepackage[normalem]{ulem}
\usepackage[mathlines]{lineno}
\usepackage[section]{placeins}
\usepackage[outline]{contour}
\usepackage[linesnumbered, ruled, vlined]{algorithm2e}
\usepackage[colorlinks = True, citecolor = blue, linkcolor = blue, anchorcolor=red, pdftoolbar = false, bookmarks = false]{hyperref}
\usepackage[utf8]{inputenc}
\usepackage{orcidlink}
\usepackage[symbol]{footmisc}

\begin{document}
\title{ Circuit Design Informed Adaptive Variational Quantum Algorithms }
\author{ Muhammad Umer \orcidlink{0000-0002-1941-1833} }
\email{umer@u.nus.edu}
\affiliation{Centre for Quantum Technologies, 3 Science Drive 2, Singapore 117543}
\author{ Dimitris G. Angelakis \orcidlink{0000-0001-6763-6060} }
\email{dimitris.angelakis@gmail.com}
\affiliation{Centre for Quantum Technologies, 3 Science Drive 2, Singapore 117543}
\affiliation{Institute for Quantum Computing and Quantum Technologies, NCSR Demokritos, Greece}
\affiliation{School of Electronics and Computer Science, University of Southampton, Southampton SO17 1BJ, UK}
\affiliation{AngelQ Quantum Computing, 531A Upper Cross Street, \#04-95 Hong Lim Complex, Singapore 051531}
\date{July 5, 2026}

\begin{abstract}
Resource-efficient computation is of central importance in the noisy intermediate-scale quantum (NISQ) era, where decoherence, gate errors, and restricted qubit connectivity severely limit the reliable execution of quantum algorithms. In this work, we demonstrate that incorporating circuit design considerations is crucial for developing resource-efficient variational quantum algorithms. By focusing on the Hadamard test circuit architecture, hardware-aware qubit connectivity, and problem-specific adaptive framework, we analyze how circuit design constraints can systematically reduce the measurement overhead associated with repeated evaluations of the candidate gate pool in adaptive algorithms. Specifically, we demonstrate reductions in the required measurement resources ranging from at least $25\%$ to as high as $50\% – 55\%$. To assess the effectiveness of our approach, we investigate the ground state problem of the nonlinear Schr\"{o}dinger equation. Overall, our work contributes to resource-friendly strategies for quantum computation and underscores that algorithmic frameworks should systematically integrate circuit design constraints with hardware-aware and problem-specific structures to enhance the practical feasibility of quantum devices in the NISQ era.
\end{abstract}
\maketitle

\section{Introduction}
\label{Sec:Introduction}

\par Quantum computation has gained significant attention in recent decades, driven by its promise to surpass the capabilities of even the most advanced supercomputers in solving problems that remain beyond the practical reach of classical methods. To exploit noisy intermediate-scale quantum (NISQ) hardware \cite{Preskill2018}, a diverse range of algorithms has been developed, including variational quantum algorithms (VQAs) \cite{Peruzzo2014, Kandala2017, Cerezo2021}, which constitute a prominent class of hybrid quantum-classical approaches with applications in quantum chemistry \cite{Peruzzo2014, Kandala2017}, computational fluid dynamics \cite{Lubasch2020, Jaksch2023}, quantum dynamics \cite{Cirstoiu2020, Lin2021, Linteau2024}, combinatorial optimization \cite{Farhi2014, Tan2021}, differential equations \cite{Bengoechea2024, Pool2024, Siegl2025, Setty2025}, and finance \cite{Huber2024, Sarma2024}. VQAs utilize a quantum device to construct approximate solutions using a variational ansatz, whose parameters are optimized on a classical computer to minimize a cost function that encodes the problem of interest \cite{Cerezo2021}.

\par Variational algorithms often involve quantum circuits that suffer from a large depth, posing significant challenges for their implementation on NISQ devices. To address this, low-depth implementations are actively pursued \cite{Grimsley2019, Sim2021, Bennakhi2024, Chan2024, Tserkis2025, Mastorakis2025}, including adaptive approaches \cite{Grimsley2019, Tang2021, Zhang2021, Yordanov2021, Zhu2022, Anastasiou2024, Feniou2025, Wu2025} that incrementally construct hardware-aware and/or problem-specific variational ans\"{a}tze by selecting gates/operators from a predefined candidate pool. These adaptive approaches involve a cost- or gradient-based selection rule in which, at each iteration, the algorithm chooses either the gate that yields the largest reduction in the cost function \cite{Feniou2025} or the gate that produces the largest cost function gradient \cite{Grimsley2019, Tang2021, Zhang2021, Yordanov2021, Zhu2022, Anastasiou2024, Wu2025}, thereby identifying the most effective route to an expressive ansatz. The pool may comprise fermionic excitations \cite{Grimsley2019}, Pauli strings \cite{Tang2021}, hardware-native gates \cite{Wu2025}, or problem-informed operators \cite{Zhu2022}, thereby balancing expressibility, trainability, and implementability on NISQ devices. 

\par In this work, we present an adaptive variational framework that, beyond incorporating hardware-aware gates, also accounts for constraints arising from the underlying circuit design, which, in turn, reduces the measurement overhead typically associated with adaptive algorithms. Specifically, by focusing on the Hadamard test (HT) design of quantum circuits, we incorporate the structural constraints associated with the Hadamard test construction \cite{Mastorakis2025} in the formulation of the gate pool. As a result of these circuit design constraints, both the size of the candidate gate pool and the measurement overhead associated with its repeated evaluations are substantially reduced. Our analysis shows that circuit design constraints reduce the scaling of the pool size, with increasing system size, from linear (quadratic) to constant (linear) in the first iteration of the adaptive algorithm, while in successive iterations the pool size grows only gradually and remains below $75\%$ of its maximum possible value. We adopt this adaptive framework to solve the ground state problem of the nonlinear Schr\"{o}dinger equation. Our results demonstrate that adaptive algorithms with design constraints yield low-depth expressive ans\"{a}tze.

\par The rest of the article is organized as follows. In section \ref{Sec:Adaptive_Scheme}, we present an adaptive framework for the construction of ans\"{a}tze. Here, we outline the constraints governing the gate selection procedure and the placement of the selected gates within adaptive ans\"{a}tze. In section \ref{Sec:NLSE}, we adopt this framework to investigate the ground state problem of the nonlinear Schr\"{o}dinger equation with varying strength of nonlinearity. Here, we examine the extent to which the trial state generated by the adaptive algorithms converges to the exact ground state. In section \ref{Sec:Resource_Comparison}, we analyze the reduction in required measurement resources relative to existing adaptive methods. The article is concluded in section \ref{Sec:Summary}.

\section{ Hadamard Test Constrained Adaptive Algorithms }
\label{Sec:Adaptive_Scheme}

\par Here, we first review the Hadamard test constraints in layered ans\"{a}tze \cite{Mastorakis2025} and discuss the resulting circuits in section \ref{Sec:Review_HTC_Ansatz}. These constraints provide the basis for the gate pool discussed in section \ref{Sec:Dynamical_Operator_Pool}, whose composition changes at each iteration of the adaptive algorithms. In section \ref{Sec:Adaptive_Algorithm}, we present the adaptive framework and discuss its various components. In particular, we describe the iterative gate selection procedure underlying the ansatz construction, through which the trial state is progressively refined.

\subsection{ Hadamard Test Constraints }
\label{Sec:Review_HTC_Ansatz}

\par
\begin{figure}[tb]
\centering
\includegraphics[clip, trim=0.42cm 0.0cm 0.6cm 0.0cm, width = 0.995\linewidth, height = 0.85\linewidth, angle=0]{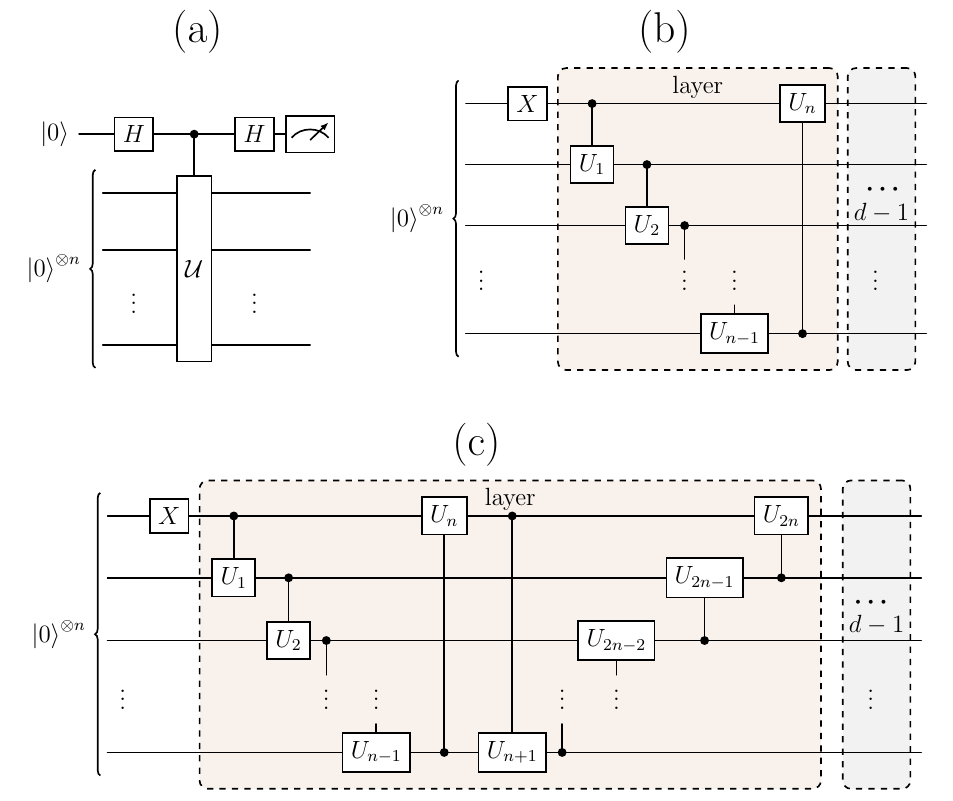}
\caption{ (a) Hadamard test and (b, c) HT constrained ans\"{a}tze. Here, $n$ is the number of qubits in the register, $d$ is the number of layers, and $U_{i} = U(\lambda_{i})$ is a controlled parameterized gate. In the construction of ans\"{a}tze, $X$ gate at the beginning of circuits is controlled by the ancilla qubit, thereby establishing ancilla$-$register correlations. Panel (b) and (c) illustrate two distinct layouts.  \vspace{-0.2cm} }
\label{fig:htc_ansatz}
\end{figure}

 The Hadamard test, shown in figure \ref{fig:htc_ansatz}a, is a circuit design strategy that is widely used in a range of quantum algorithms \cite{Jaksch2023, Sarma2024}. Recently, design constraints that leverage the logical composition of HT circuits were investigated in Ref. \cite{Mastorakis2025} and shown to facilitate low-depth circuit constructions. A key observation is that, when all qubits are initialized in the $\ket{0}$ state, conditional gates whose control qubits reside within the register can be implemented without ancilla control. Accordingly, a class of ans\"{a}tze can be composed using only conditional parameterized gates acting within the quantum register, as illustrated in figures \ref{fig:htc_ansatz}b$-$\ref{fig:htc_ansatz}c. A single ancilla-controlled $X$ gate is applied to one of the qubits in the register to establish ancilla$-$register correlations, and this remains the only required ancilla$-$register interaction in this ansatz architecture. Subsequently, all layers consist only of intra-register controlled rotations, which both generate entanglement within the register and introduce the variational degrees of freedom. A defining design constraint is that control qubits are restricted to those that have previously served as targets, thereby enforcing a HT compatible circuit structure. It is worth emphasizing that, although these insights enable low-depth circuit constructions, as shown in Ref. \cite{Mastorakis2025} for nonlinear Burgers\textquotesingle{} dynamics, an identical predefined layer sequence might not be an optimal gate ordering for a given problem. Consequently, it is expected that an adaptive framework may generate an unstructured gate sequence tailored to the specific problem, potentially further reducing circuit depth.

\subsection{ Constrained Gate Pool }
\label{Sec:Dynamical_Operator_Pool}

\par Here, we focus on constraints that govern the selection of gates constituting the candidate pool $\mathcal{O}_{\rm pool}$. Specifically, gate types are fixed a priori, for example, rotations $R_{x}$, $R_{y}$, and $R_{z}$, together with controlled rotations $CR_{x}$, $CR_{y}$, and $CR_{z}$. Constraints then primarily determine the qubit pairs on which controlled operations may act, thereby limiting the size of $\mathcal{O}_{\rm pool}$. These constraints stem from (i) the qubit connectivity of the hardware platform, (ii) the HT compatible structure induced by the gates chosen in preceding iterations, and (iii) redundant gates on the same qubit(s). In what follows, we discuss each of these constraints in detail. 

\begin{figure}[t!]
\centering
\includegraphics[clip, trim=0.0cm 0.0cm 0.0cm 0.0cm, width = 0.99\linewidth, height = 0.52\linewidth, angle=0]{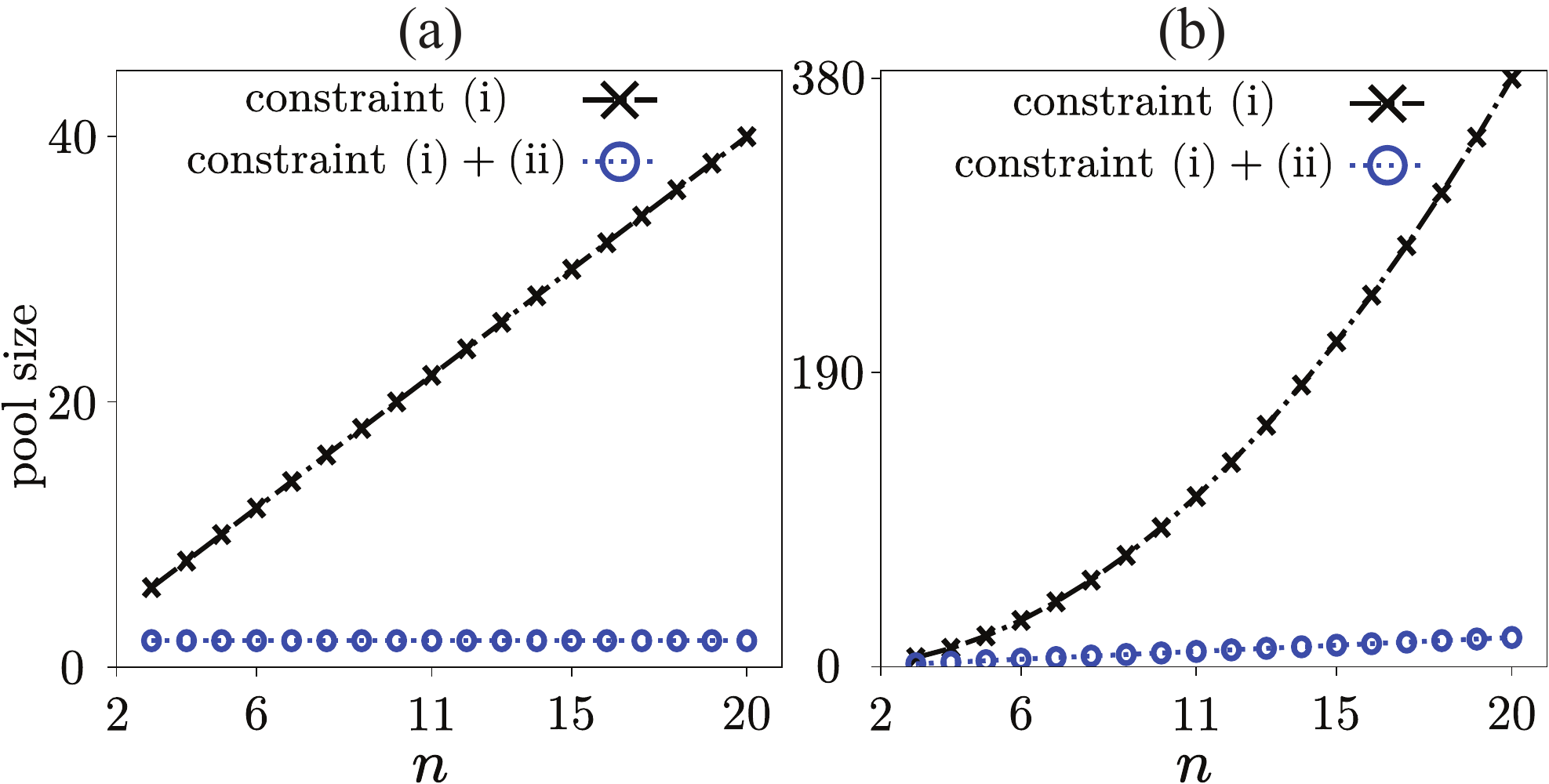}
\caption{ Scaling behavior of the gate pool $\mathcal{O}_{\rm pool}$ at the first iteration of adaptive frameworks. (a) and (b) correspond to the nearest-neighbor connectivity in a ring topology and all-to-all connectivity, respectively. Here, constraints (i) and (ii) are imposed by qubit connectivity and Hadamard test circuit design, respectively. \vspace{-0.2cm} }
\label{fig:pool_scaling}
\end{figure}

\par First, the qubit connectivity of hardware platforms plays a central role in the multi-qubit gate decomposition and, accordingly, serves as an important parameter governing the gate pool $\mathcal{O}_{\rm pool}$ in a hardware-aware adaptive framework \cite{Tang2021, Feniou2025, Wu2025}. On the one hand, limited qubit connectivity restricts the set of qubit pairs for direct realization of two-qubit operations and generally requires deeper compiled circuits for the implementation of distant-qubit interactions \cite{Herbert2018}; on the other hand, all-to-all connectivity permits a broader set of qubit pairs and, consequently, shallower implementations of the same interactions. Here, the set of connected qubit pairs determines the size of $\mathcal{O}_{\rm pool}$, which in turn dictates the number of circuit evaluations required to assess the gate selection criterion, and thereby constitutes a major bottleneck for hardware-aware adaptive frameworks \cite{Tang2021, Feniou2025, Wu2025}. The black curve in figure \ref{fig:pool_scaling}a (\ref{fig:pool_scaling}b) illustrates the linear (quadratic) scaling behavior of $\mathcal{O}_{\rm pool}$ in the first iteration of adaptive algorithms. In figures \ref{fig:pool_scaling}a and \ref{fig:pool_scaling}b, we consider a single type of two-qubit gate with nearest-neighbor connectivity in a ring topology and with all-to-all connectivity, respectively. 

\par The second constraint arises from circuit design, i.e., the Hadamard test structure \cite{Mastorakis2025}. This prunes $\mathcal{O}_{\rm pool}$ by restricting the admissible control qubits to those that have already been involved in the ansatz up to the current iteration of the adaptive algorithm. Specifically, a controlled operation acting on a pair of qubits is admitted to $\mathcal{O}_{\rm pool}$ only if its control has previously been acted upon through single- or two-qubit gates. This mitigates the measurement overhead associated with evaluating a large set of candidate pairs. Moreover, this constraint enables a key mechanism that renders $\mathcal{O}_{\rm pool}$ dynamical: at each iteration, the admissible candidate set is updated based on the gates selected in the preceding step, systematically introducing new gates into $\mathcal{O}_{\rm pool}$ in accordance with the most recent selection. The blue curve in figure \ref{fig:pool_scaling}a (\ref{fig:pool_scaling}b) illustrates constant (linear) scaling of $\mathcal{O}_{\rm pool}$ after pruning the hardware-aware candidate pool according to HT constraints in the first iteration of adaptive frameworks. Here, constant and linear scaling behaviors correspond to nearest-neighbor connectivity in a ring topology and all-to-all connectivity, respectively. The reduction in pool size scaling, from linear (quadratic) to constant (linear), already demonstrates a substantial decrease in the measurement resources required for evaluating the gate selection criterion.

\begin{figure}[t!]
\centering
\includegraphics[clip, trim=0.0cm 0.0cm 0.0cm 0.0cm, width = 0.99\linewidth, height = 0.38\linewidth, angle=0]{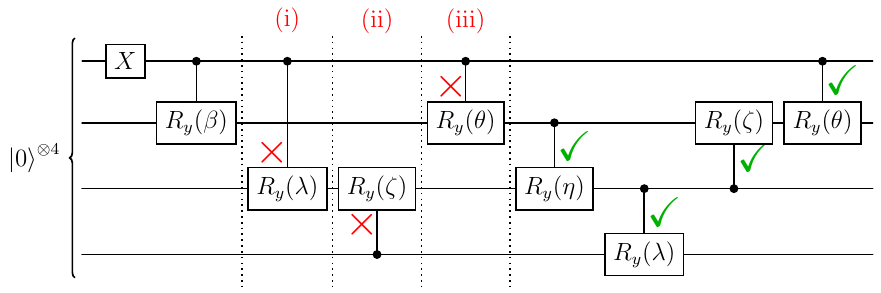}
\caption{Constraints embedded in the gate pool $\mathcal{O}_{\rm pool}$. Here, (i)–(iii) highlight three classes of constraints, namely: (i) qubit connectivity constraint of the hardware platform, (ii) circuit design, i.e., the Hadamard test constraint, and (iii) non-redundancy constraints that prevent immediate re-selection of gates acting on the same qubit(s) by temporarily removing them from the candidate pool. \vspace{-0.2cm}}
\label{fig:operator_constraints}
\end{figure}

\par Third, a hardware-aware scheme imposes a non-redundancy constraint. This allows exclusion of gates that would induce repeated applications of the same primitive operation on identical qubit(s). For example, once a controlled rotation is appended to the ansatz, we remove it from $\mathcal{O}_{\rm pool}$ to prevent immediate re-selection, and reinstate it only after another gate acts on at least one of the control$-$target qubits. 

\par Figure \ref{fig:operator_constraints} illustrates how the aforementioned constraints influence the construction of ans\"{a}tze. First, annotation (i) in figure \ref{fig:operator_constraints} highlights hardware constraint where controlled operations between non-adjacent qubits are disallowed on a device with nearest-neighbor connectivity; consequently, gates acting on distant qubit pairs are excluded from $\mathcal{O}_{\rm pool}$. Second, annotation (ii) in figure \ref{fig:operator_constraints} illustrates HT compatibility constraint. It emphasizes that operations whose control qubit has not been previously acted upon are not allowed and are therefore excluded from $\mathcal{O}_{\rm pool}$. Finally, annotation (iii) in figure \ref{fig:operator_constraints} underscores the non-redundancy constraint, which prohibits the immediate reselection of gates acting on the same qubit(s). Accordingly, such gates are temporarily excluded from $\mathcal{O}_{\rm pool}$ and are reinstated only after an intervening operation has acted on at least one of the corresponding qubits. Collectively, these constraints reduce the number of candidate gates evaluated at each iteration while enforcing a hardware-aware HT ansatz construction.

\subsection{ Adaptive Algorithms }
\label{Sec:Adaptive_Algorithm}

\par Having described the set of constraints that govern the gate pool and the resulting ansatz design, we now present adaptive algorithms, that follow Refs. \cite{Grimsley2019, Feniou2025}, adopted to generate variational ans\"{a}tze. 

\begin{algorithm}[hbt]
\caption{ Adaptive Algorithms } \label{Algo:Adaptive}
evaluate $\mathcal{C}_{0}$ with only $X$ gate in the ansatz \\
\textbf{Input:} primitive gate set, qubit connectivity, and $r = 0$ \\
\For{ $k = 1, \cdots, \mathcal{N}_{\rm iteration}$ }{
    compose/update $\mathcal{O}_{\rm pool}$\\
    \For{${\rm each}~U(\lambda_{p}) \in \mathcal{O}_{\rm pool}$ }{
        evaluate cost- or gradient-based criterion
        }
    evaluate $\epsilon_{k}$, i.e.,  convergence indicator \\
    \eIf{$\epsilon_{k} < \epsilon_{\rm threshold}$}{
        terminate }{
        temporarily append top $m$ gates in ansatz and evaluate $\mathcal{C}_{k}^{'}$\\
        \eIf{$\mathcal{C}_{k}^{'} \le \mathcal{C}_{k-1}$ {\rm or} $r = 2$ {\rm and} $\mathcal{C}_{k}^{'} \le f^{'}\mathcal{C}_{k-1}$ {\rm or} $r > 2$}{retain $m$ gates in ansatz and reset $r = 0$}{
        configure $r = r + 1$
        }
        optimize all variational parameters
        }
    }
\Return{\rm sequence of gates and corresponding optimized variational parameters}
\end{algorithm}

\par Here, we discuss several key steps of the adaptive framework presented in Algorithm \ref{Algo:Adaptive}, where each iteration $k$ is carried out as follows. Given an updated $\mathcal{O}_{\rm pool}$, for each candidate gate $U(\lambda_{p})$, we evaluate either the minimum value of the cost function $\underset{\lambda_{p}}{\min}[\mathcal{C}_{k}(\boldsymbol{\lambda}_{k-1}, \lambda_{p})]$ and corresponding parameter value $\underset{\lambda_{p}}{\arg\min}[\mathcal{C}_{k}(\boldsymbol{\lambda}_{k-1}, \lambda_{p})]$  or the cost function gradient $\frac{\partial\mathcal{C}_{k}(\boldsymbol{\lambda}_{k-1}, \lambda_{p})}{\partial\lambda_{p}}\big\vert_{\lambda_{p} = 0}$ depending on the selection criterion. Here, $\boldsymbol{\lambda}_{k-1}$ is the set of parameters optimized in the $(k-1){\rm th}$ iteration, and we evaluate cost function  $\mathcal{C}_{k}(\boldsymbol{\lambda}_{k-1}, \lambda_{p})$ by utilizing the procedure discussed in Ref. \cite{Umer2025opt}. This allows us to evaluate $\epsilon_{k}$, which serves as a measure of the average variation induced by candidate gates and, hence, of the extent to which the current ansatz remains insufficiently expressive with respect to the span generated by $\mathcal{O}_{\rm pool}$ \cite{Grimsley2019, Tang2021, Zhang2021, Yordanov2021, Zhu2022, Anastasiou2024}. For the minimum cost criterion, we define $\epsilon_{k} = \sum_{p}\bigl[\mathcal{C}_{k-1} - \underset{\lambda_{p}}{\min}[\mathcal{C}_{k}(\boldsymbol{\lambda}_{k-1}, \lambda_{p})]\bigl]$, namely, the sum of residuals of the cost function value attained in the previous iteration and the minimum values achieved by each candidate gate. Because each difference is non-negative, $\epsilon_{k}$ quantifies the collective variation induced by candidate gates. For the gradient criterion, on the contrary, we define $\epsilon_{k} = \sqrt{\sum_{p}\bigl[\frac{\partial{\mathcal{C}_{k}(\boldsymbol{\lambda}_{k-1}, \lambda_{p})}}{\partial{\lambda_{p}}}\big|_{\lambda_{p} = 0}\bigl]^{2}~}$ as the norm of the cost function gradient for all candidate gates, consistent with the criterion adopted in previous studies \cite{Grimsley2019, Zhu2022}. At this stage, if $\epsilon_{k} < \epsilon_{\rm threshold}$, the adaptive routine is terminated; otherwise, it is continued. 

\par If $\epsilon_{k} \ge \epsilon_{\rm threshold}$, further analysis is performed to determine whether to grow ansatz by adding gates to the variational circuit or to retain it in its current form. In this regard, we select $m$ candidate gates according to the largest variations they \emph{individually} induce in the cost function. These $m$ gates are then temporarily added to the ansatz, after which cost function value $\mathcal{C}_{k}^{'}$ is evaluated. If $\mathcal{C}_{k}^{'} \le \mathcal{C}_{k-1}$, $m$ gates are permanently retained in the ansatz, and all variational parameters are subsequently optimized using the sequential grid-based explicit optimization (SGEO) routine \cite{Umer2025opt}. Otherwise, the following procedure is adopted. At first, $m$ gates are not retained, and only the existing variational parameters are re-optimized. If $\mathcal{C}_{k}^{'} \le \mathcal{C}_{k-1}$ remains unsatisfied in the subsequent iteration, $m$ gates are retained only if $\mathcal{C}_{k}^{'} \le f^{'}\mathcal{C}_{k-1}$ with $f^{'} > 1$ is satisfied, thereby relaxing the acceptance criterion. If this condition is likewise not met, all variational parameters are re-optimized once more, after which $m$ gates are retained in the ansatz regardless of acceptance criterion. 

\par A few comments are in order. For the maximum gradient criterion, the condition $\mathcal{C}^{'}_{k} \le \mathcal{C}_{k-1}$ is always satisfied. Consequently, the adaptive framework proceeds in a manner analogous to ADAPT-VQE \cite{Grimsley2019, Wu2025}, with the ansatz growing by $m$ gates in each iteration. For the minimum cost criterion, the condition $\mathcal{C}^{'}_{k} \le \mathcal{C}_{k-1}$ may fail to hold, thereby triggering the execution of the relaxed acceptance routine mentioned above. This relaxed acceptance routine is designed to avoid the need for rollback procedures \cite{Wu2025}, wherein gates are removed from the ansatz if the cost function value increases rapidly in subsequent iterations. Although more sophisticated strategies may be developed in this context, we do not pursue them in the present work. 

\par In what follows, we employ these adaptive frameworks to analyze the ground state of the nonlinear Schr\"{o}dinger equation and assess the resource savings achieved by incorporating circuit design constraints, relative to the unconstrained case.

\section{Nonlinear Schr\"{o}dinger Equation}
\label{Sec:NLSE}

\par The nonlinear Schr\"{o}dinger equation (NLSE) is a one-dimensional time-independent equation. It helps model various nonlinear phenomena in Bose-Einstein condensates \cite{Gross1961, *Pitaevskii1961, Dalfovo1999} and optics \cite{Agrawal2000}, to name a few. In dimensionless form, the NLSE with quadratic potential and interaction terms is given as 
\begin{eqnarray}\begin{aligned}
\big[-\frac{1}{2}\frac{d^{2}}{d{x}^{2}} + V_{0}\big(x - x_{0}\bigl)^{2} + g\vert{u(x)}\vert^{2}\big] {u(x)} = E {u(x)}\;.~~~
\label{EQ:Schrodinger}
\end{aligned}\end{eqnarray}
Here, $u(x)$, with $x$ being spatial coordinates, represents a normalized single real-valued function defined over the interval $[a, b]$. The parameter $g$ denotes the strength of nonlinearity, and $V_{0}$ is the depth of the quadratic potential $V(x)$ centered around $x_{0} = (b-a)/2$. Furthermore, we consider periodic boundary conditions such that $u(b) = u(a)$ and $V(b) = V(a)$. In this work, we adopt $V_{0} = 10^{3}$ and vary the nonlinearity strength $g$ to investigate distinct target ground states across nonlinear regimes. Additionally, we employ the classical method of imaginary-time evolution \cite{Edwards1995} to benchmark the variational results.

\par For ground state problem of the NLSE on $n$ qubits, we discretize the spatial domain into $2^n$ grid points and consider the finite-difference method (refer to Refs. \cite{Lubasch2020, Umer2025opt} for details). We define $\psi_{k} = \sqrt{\delta}u(x_{k})$ with $\delta$ being the separation between adjacent grid points such that the normalization condition $1 = \delta\sum_{k = 0}^{N-1} \vert{u(x_k)}\vert^{2} = \sum_{k = 0}^{N-1} \vert{\psi_{k}}\vert^{2}$ is satisfied. The cost function is then defined as the sum of interaction, kinetic, and potential energy, $\langle\langle{E}\rangle\rangle = \langle\langle{E_{I}}\rangle\rangle + \langle\langle{E_{K}}\rangle\rangle + \langle\langle{E_{P}}\rangle\rangle$ \cite{Lubasch2020}, where
\begin{eqnarray}\begin{aligned}
\label{EQ:Cost_Functions}
\langle\langle{E_{P}}\rangle\rangle &= \sum_{k = 0}^{N-1}~\vert\psi_{k}\vert^{2}V_{k}\;,~~~~ \langle\langle{E_{I}}\rangle\rangle = \sum_{k = 0}^{N-1}~\frac{g}{\delta}\vert\psi_{k}\vert^{4}\;,\\
\langle\langle{E_{K}}\rangle\rangle &= \frac{1}{2\delta^{2}}\bigl[ 2 - \sum_{k = 0}^{N-1}\big(\psi_{k}\psi_{k+1} + \psi_{k}\psi_{k-1}\big) \bigl]\;,
\end{aligned}\end{eqnarray}
and the minimum value of the cost function $\langle\langle{E}\rangle\rangle$ represents the ground state solution. Each term of the cost function requires a separate quantum circuit \cite{Umer2025opt, Mastorakis2025}. It is worth emphasizing that simulating a problem instance defined on $n$ qubits requires a total of $3n + 1$ qubits \cite{Umer2025opt}, which in turn limits the system sizes that can be investigated numerically.

\subsection{Results: Layered Hadamard Test Ans\"{a}tze}
\label{Sec:Fixed_HTC}

\begin{figure}[t!]
\centering
\includegraphics[clip, trim=0.0cm 0.0cm 0.0cm 0.0cm, width = 0.99\linewidth, height = 1.05\linewidth, angle=0]{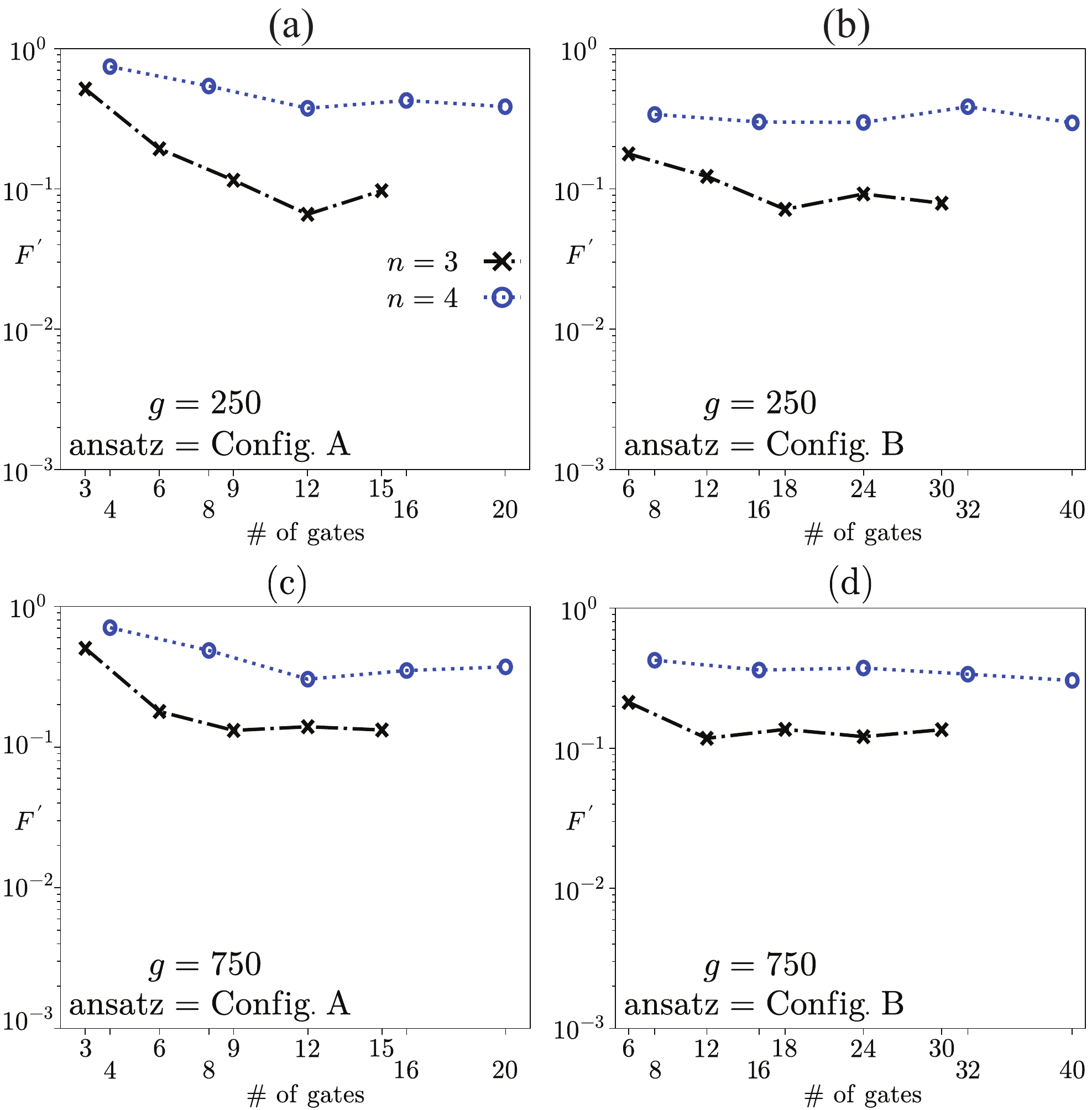}
\caption{ Behavior of state infidelity as a function of number of gates in the layered HT ans\"{a}tze. State infidelity quantifies the deviation of the variational state from the target classical ground state. Results are shown for $g = 250$ in panels (a, b) and for $g = 750$ in panels (c, d). In addition, layer configuration A and B is adopted in panels (a, c) and (b, d), respectively. \vspace{-0.2cm} }
\label{fig:nlse_fixed}
\end{figure}

\par Before analyzing the adaptive framework, we first examine the expressivity of ans\"{a}tze composed of layers that satisfy the HT constraints. In this context, we consider two layouts: configuration A shown in figure \ref{fig:htc_ansatz}b and configuration B shown in figure \ref{fig:htc_ansatz}c, where every $CU_{i}$ gate is replaced by $CR_{y}(\lambda_{i})$. Under these settings, we assess the ability of layered HT ans\"{a}tze to capture the ground state of the NLSE for nonlinearity strengths $g = 250$ and $g = 750$, and present the results in figure \ref{fig:nlse_fixed}. Although the cost function is defined in terms of the energy expectation value, here we instead report the state infidelity $F^{'} = 1- \vert\langle\Psi_{\rm target}\vert\Psi_{\rm var}\rangle\vert^{2}$ with respect to the exact classical ground state, as this provides a direct measure of the quality of variational solutions. State infidelity $F^{'}$ is plotted in figure \ref{fig:nlse_fixed} as a function of the number of gates, where the latter is increased by adding layers to the ansatz. Specifically, we vary the number of layers from $1$ to $5$ and observe that resulting variational states attain at most only $90\% - 93\%$ fidelity for $n = 3$ which drops to $50\% - 70\%$ for $n = 4$. Figure \ref{fig:nlse_fixed} illustrates that state fidelity does not increase as we increase the number of gates, demonstrating that layered HT ans\"{a}tze, depicted in figures \ref{fig:htc_ansatz}b-\ref{fig:htc_ansatz}c, are insufficiently expressive to accurately represent the ground state of the NLSE in these regimes of nonlinearity. 

\par It is important to highlight a few key points. Here, we consider only two specific layouts; however, an arbitrarily large number of layouts can, in principle, be constructed within the class of layered HT ans\"{a}tze. Therefore, we do not imply that layered ans\"{a}tze are intrinsically incapable of achieving higher state fidelity. For example, two such ans\"{a}tze, which do not satisfy the HT constraints, were studied in Refs. \cite{Umer2025opt} and \cite{Umer2025} and were shown to achieve higher fidelities for the same problem instances. The present results, however, support the argument that repeated application of the same layer does not necessarily yield an optimal gate sequence, while analyzing an arbitrarily large number of layer layouts remains highly nontrivial and computationally expensive.

\subsection{Results: Adaptive Hadamard Test Ans\"{a}tze}
\label{Sec:Adaptive_HTC}

\begin{figure}[t!]
\centering
\includegraphics[clip, trim=0.0cm 1.8cm 0.0cm 0.0cm, width = 0.99\linewidth, height = 1.20\linewidth, angle=0]{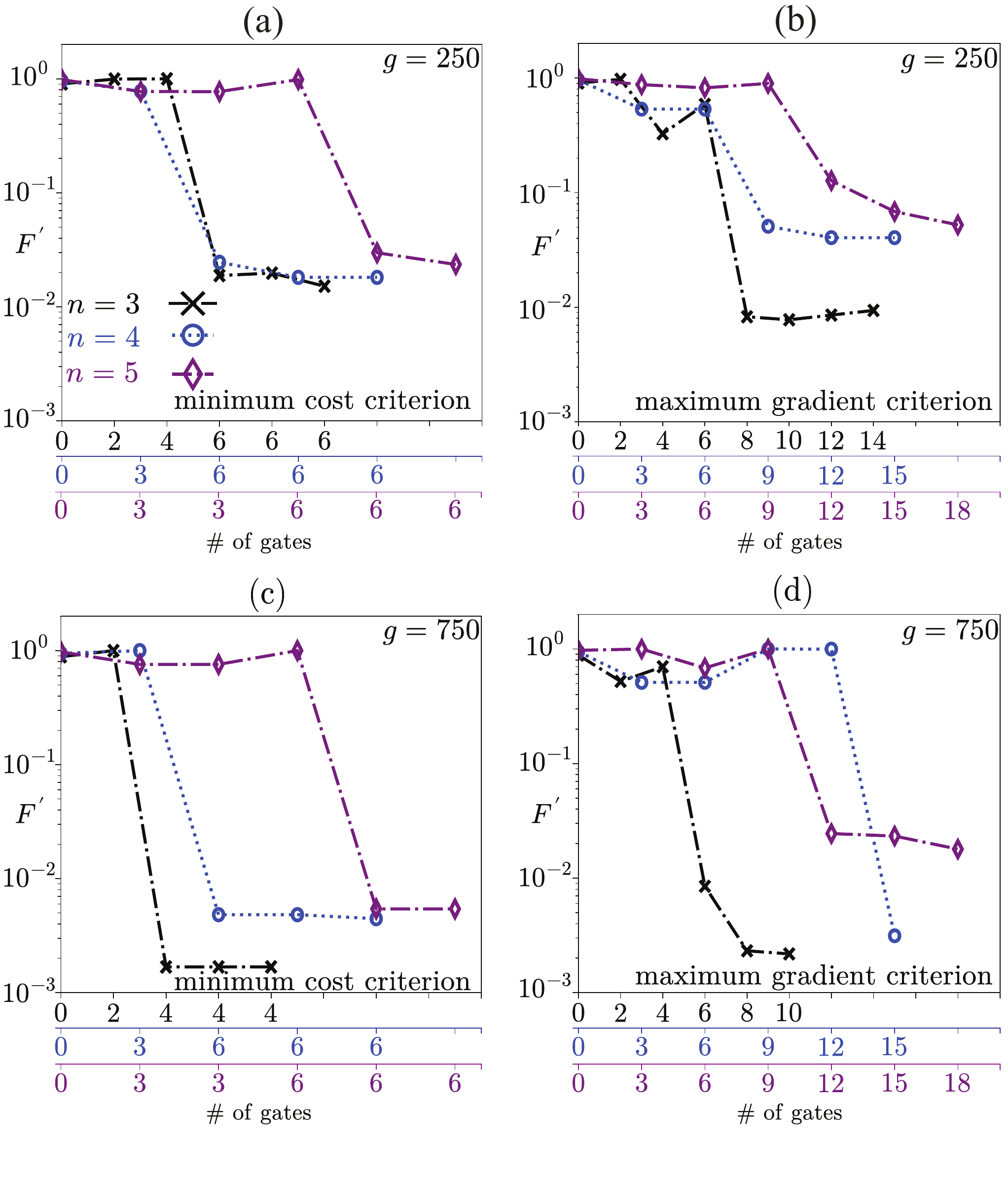}
\caption{ Behavior of the state infidelity as a function of number of gates in adaptive HT ans\"{a}tze. Results are shown for $g = 250$ in panels (a, b) and for $g = 750$ in panels (c, d). In addition, minimum cost in (a, c) and maximum gradient in (b, d) are two criteria for gate selection. Here, each curve is associated with its own horizontal axis, indicated by the corresponding color: black for $n = 3$, blue for $n = 4$, and purple for $n = 5$. \vspace{-0.2cm} }
\label{fig:nlse_adaptive}
\end{figure}

\par We now turn to adaptive framework to construct HT ans\"{a}tze. In this regard, we consider $n = 3$, $n = 4$, and $n = 5$ qubit systems, corresponding to $2^n$ grid points. Here, we consider the primitive gate set $\{R_{y},~CR_{y}\}$, where $CR_{y}$ gates act only between the nearest-neighbor qubits arranged in a circular geometry. We consider two distinct gate selection strategies within the adaptive framework. In the first scheme, new gates are selected according to the minimum value of the cost function, $\min\limits_{p,~\lambda_{p}}[\mathcal{C}_{k}(\boldsymbol{\lambda}_{k-1}, \lambda_{p})]$, and we refer to this as the minimum cost criterion \cite{Feniou2025}. In the second scheme, gate selection is based on the largest gradient of the cost function at zero parameter value $\max\limits_{p}\bigl[\frac{\partial\mathcal{C}_{k}(\boldsymbol{\lambda}_{k-1}, \lambda_{p})}{\partial\lambda_{p}}\big\vert_{\lambda_{p}=0}\bigl]$, following the strategy adopted in ADAPT-VQE \cite{Grimsley2019} and we refer to this as the maximum gradient criterion. Additionally, in both cases, we append more than one gate to the ansatz during each iteration. Specifically, we fix $m = 2$ for $n = 3$, and $m = 3$ for $n = 4, 5$. Throughout the following analysis, we employ the constrained gate pool introduced in section \ref{Sec:Dynamical_Operator_Pool}. 

\par Results obtained utilizing the constrained adaptive algorithms are presented in figure \ref{fig:nlse_adaptive}. First, we consider the maximum gradient criterion for generating HT ans\"{a}tze. Figures \ref{fig:nlse_adaptive}b and \ref{fig:nlse_adaptive}d highlight that the adaptive algorithm generates HT ans\"{a}tze yielding trial states with fidelity exceeding $95\%$ for $g = 250$ and $99\%$ for $g = 750$, relative to classical results. 
Second, we consider the minimum cost criterion where we choose $f^{'} = 1.35$ for a relaxed acceptance criterion for gate selection. Figures \ref{fig:nlse_adaptive}a and \ref{fig:nlse_adaptive}c demonstrate that, under the minimum cost criterion, the adaptive algorithm constructs ans\"{a}tze that prepare trial states with fidelities exceeding $98\%$ for $g = 250$ and $99\%$ for $g = 750$, relative to classical results. These results highlight that the constrained gate pool contains the essential gates required to construct expressive, low-depth ans\"{a}tze. 

\par
\begin{figure}[t!]
\centering
\includegraphics[clip, trim=0.0cm 0.0cm 0.0cm 0.0cm, width = 0.99\linewidth, height = 0.98\linewidth, angle=0]{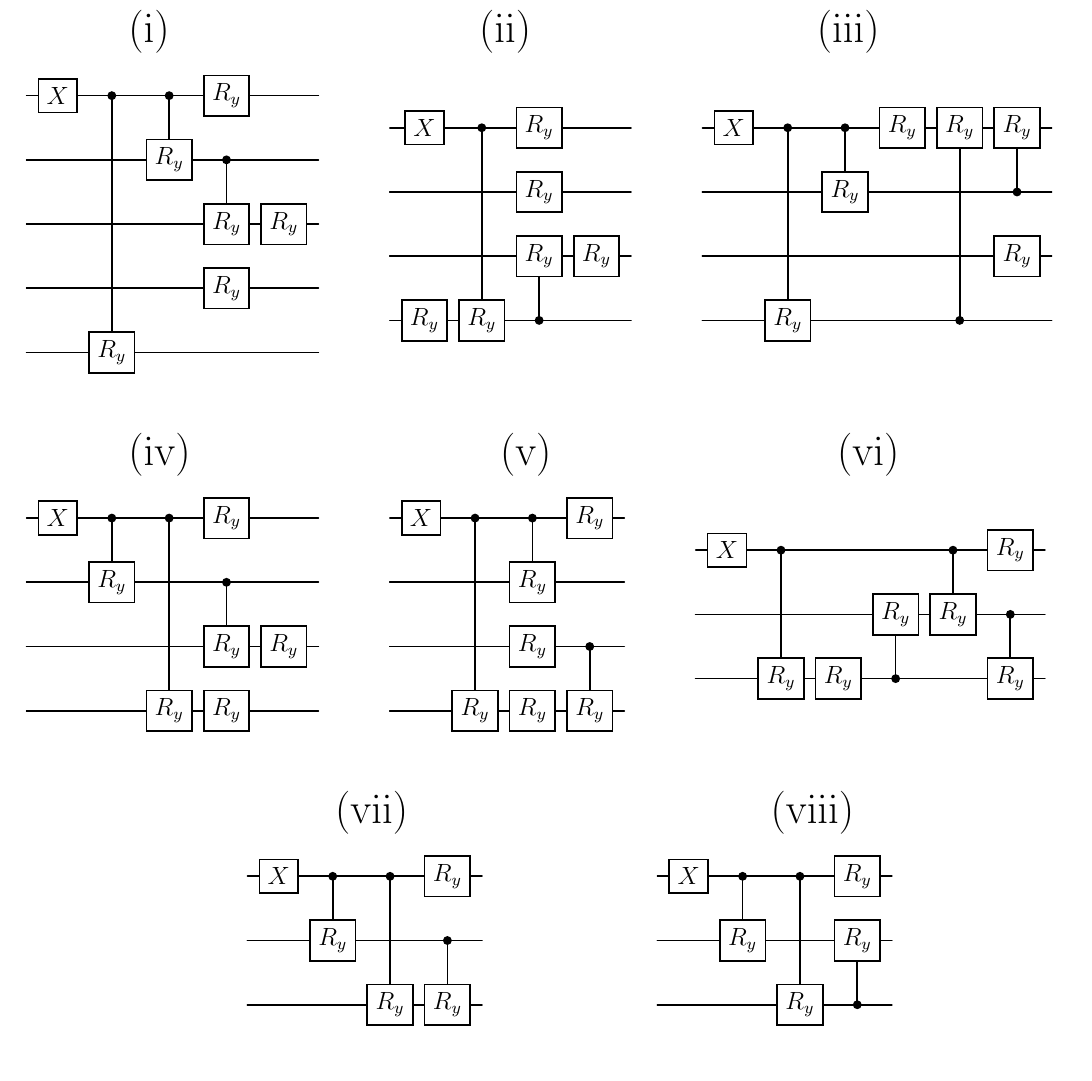}
\caption{ Adaptive HT ans\"{a}tze. Panels are organized according to both system size and nonlinear interaction strength. Here, $n = 5$ for panel (i), $n = 4$ in panels (ii$-$v), and $n = 3$ in panels (vi$-$viii). Additionally, $g = 250,~500,~750,$ and $1000$ correspond respectively to panels (i, ii, vi), (iii, vii), (iv, viii), and (v, viii).  \vspace{-0.2cm} }
\label{fig:nlse_adaptive_ansatzes}
\end{figure}

 Several aspects merit further discussion. Each successive data point in figure \ref{fig:nlse_adaptive} corresponds to a subsequent iteration, while the horizontal axis represents the number of gates in HT ans\"{a}tze. In figures \ref{fig:nlse_adaptive}b and \ref{fig:nlse_adaptive}d, $m$ gates are added to grow ansatz in each iteration under the maximum gradient criterion. By contrast, not every iteration leads to an increase in the number of gates under the minimum cost criterion, as depicted in figures \ref{fig:nlse_adaptive}a and \ref{fig:nlse_adaptive}c. However, re-optimization of the existing parameters without appending new gates either decreases the infidelity or leaves it unchanged. This aspect highlights the flexibility of the adaptive framework in avoiding unnecessary ansatz growth that would otherwise drastically perturb the gate sequence and lead to an excessive increase in the state infidelity. This behavior demonstrates that the adaptive framework provides a stable and systematically improving route for constructing ans\"{a}tze with an enhanced accuracy. Compared to layered ans\"{a}tze presented in figures \ref{fig:htc_ansatz}b$-$\ref{fig:htc_ansatz}c, adaptive ans\"{a}tze are more expressive, comprising an appropriate sequence of gates constructed according to the requirements of the underlying problem. Specifically, figure \ref{fig:nlse_adaptive_ansatzes} presents ans\"{a}tze obtained using the minimum cost criterion within the adaptive framework. Figure \ref{fig:nlse_adaptive_ansatzes}(i) shows the ansatz for $n = 5$ with $g = 250,~ 750$. Figures \ref{fig:nlse_adaptive_ansatzes}(ii) and \ref{fig:nlse_adaptive_ansatzes}(iv) correspond to $n = 4$ with $g = 250$ and $g =750$, respectively, while figures \ref{fig:nlse_adaptive_ansatzes}(vi) and \ref{fig:nlse_adaptive_ansatzes}(viii) show ans\"{a}tze for $n = 3$ with $g = 250$ and $g = 750$. These circuits demonstrate that the constrained adaptive framework yields shallow and compact ans\"{a}tze.

\par
\begin{figure}[t!]
\centering
\includegraphics[clip, trim=0.0cm 0.0cm 0.0cm 0.0cm, width = 0.99\linewidth, height = 0.49\linewidth, angle=0]{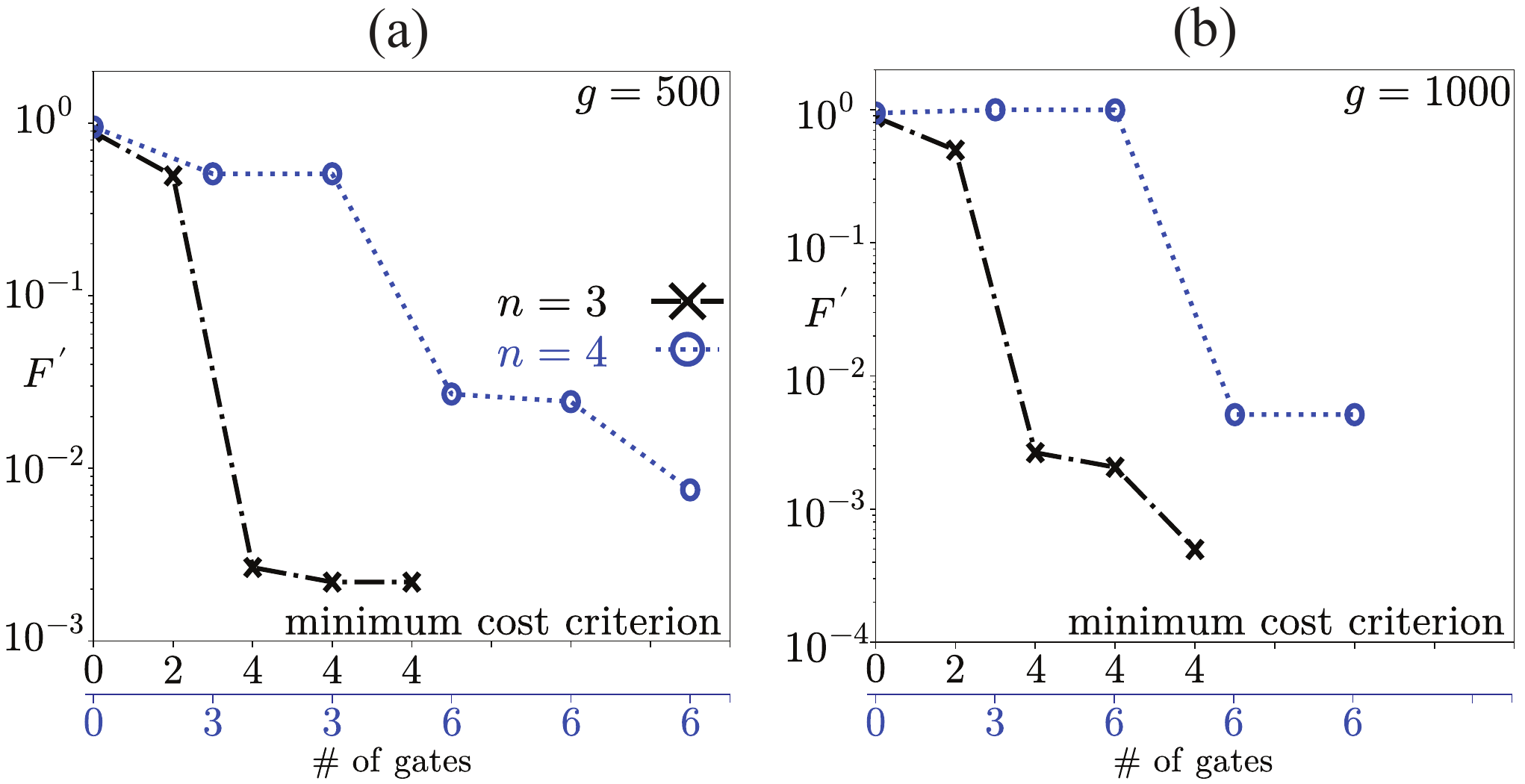}
\caption{ State infidelity as a function of the number of gates in HT ans\"{a}tze. Results are shown for $g = 500$ in panels (a) and for $g = 1000$ in panels (b), where minimum cost criterion is adopted within adaptive algorithm. Each curve is associated with its own horizontal axis, indicated by the corresponding color: black for $n = 3$ and blue for $n = 4$. \vspace{-0.2cm} }
\label{fig:nlse_adaptive_additional}
\end{figure}

 To further assess the performance of adaptive algorithm under the minimum cost criterion, we extend the analysis to additional values of the nonlinear interaction strength. Specifically, we consider $n = 3$ and $n = 4$ qubit cases at nonlinearity strengths $g = 500$ and $g = 1000$, and present the corresponding results in figure \ref{fig:nlse_adaptive_additional}. These results are fully consistent with the trends presented in figures \ref{fig:nlse_adaptive}a and \ref{fig:nlse_adaptive}c. In particular, the constrained adaptive framework continue to construct expressive ans\"{a}tze while maintaining a low gate count, as shown in figures \ref{fig:nlse_adaptive_ansatzes}(iii), \ref{fig:nlse_adaptive_ansatzes}(v),  \ref{fig:nlse_adaptive_ansatzes}(vii), and \ref{fig:nlse_adaptive_ansatzes}(viii). Overall, the results presented in this section demonstrate that incorporating circuit design constraints into the gate pool provides an effective route to constructing expressive, low-depth circuits for target ground states associated with distinct nonlinearity regimes.

\par These results may be compared with pool constructions that either do not incorporate circuit-design constraints \cite{Grimsley2019, Tang2021, Zhang2021, Yordanov2021, Zhu2022, Anastasiou2024, Feniou2025, Wu2025} or assume all-to-all qubit connectivity, as summarized below. In such alternative settings, the candidate pool would still contain the gates selected by the constrained adaptive algorithm subject to nearest-neighbor qubit connectivity, along with many additional admissible gates, and would therefore remain capable of generating similar ans\"{a}tze. Nevertheless, screening such an enlarged candidate pool would require significantly higher measurement resources.

\subsection{Measurement-Resource Reduction}
\label{Sec:Resource_Comparison}

\par In an adaptive framework, repeated execution of shallow quantum circuits to identify the most suitable gates from the pool gives rise to substantial measurement overhead, which is one of the most resource intensive aspects of this procedure. Consequently, the measurement cost is directly tied to the size of the gate pool. Reducing this overhead is therefore of central importance. As shown in section \ref{Sec:Dynamical_Operator_Pool}, incorporating circuit design constraints reduces the scaling of $\mathcal{O}_{\rm pool}$ with system size from linear (quadratic) to constant (linear) in the first iteration of adaptive frameworks. In this section, we further analyze the case of nearest-neighbor qubit connectivity and investigate how $\mathcal{O}_{\rm pool}$ size evolves over successive iterations. In this regard, in each iteration we randomly select three gates from $\mathcal{O}_{\rm pool}$ to grow the ansatz, subsequently update $\mathcal{O}_{\rm pool}$ according to the constraints discussed in section \ref{Sec:Dynamical_Operator_Pool}, and average the resulting behavior over $100$ distinct realizations. 

\begin{figure}[t!]
\centering
\includegraphics[clip, trim=0.0cm 0.0cm 0.0cm 0.0cm, width = 0.99\linewidth, height = 0.75\linewidth, angle=0]{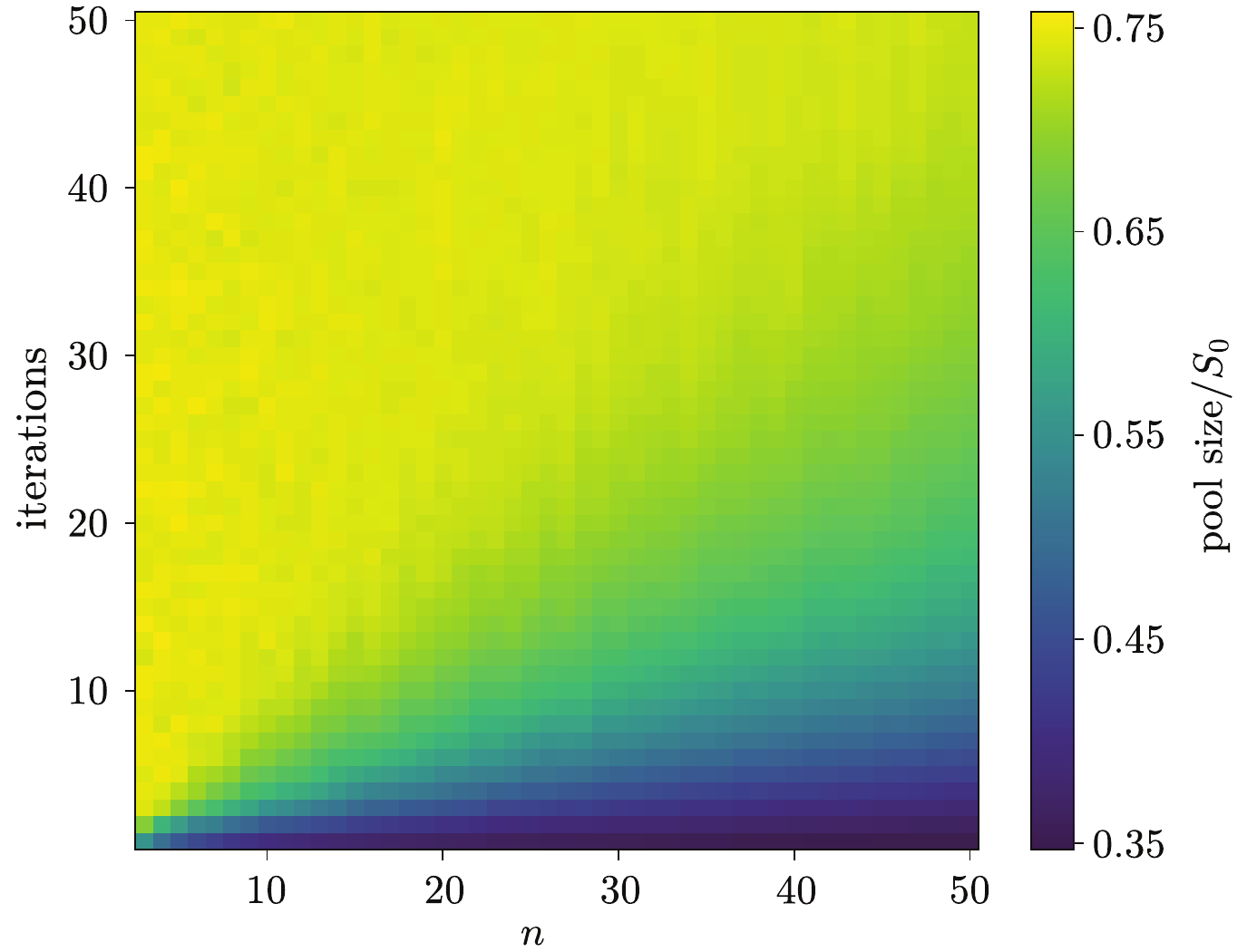}
\caption{ Fraction of the pool size relative to its maximum possible size $S_{0}$ as a function of the system size $n$ and the number of iterations. Here, the pool size is estimated based on random gate selections, without performing actual quantum simulations or adaptive ansatz construction. \vspace{-0.2cm} }
\label{fig:pool_size}
\end{figure}

\par Figure \ref{fig:pool_size} illustrates the fraction of $\mathcal{O}_{\rm pool}$ size relative to its maximum possible size $S_{0}$ as a function of $n$ and the number of iterations. This highlights that, at any given iteration and independently of $n$, $\mathcal{O}_{\rm pool}$ contains only up to $75\%$ of the gates compared to its maximum possible size $S_{0}$. Consequently, the measurement resources are reduced by at least $25\%$, with substantially larger reductions exceeding $50\%-55\%$ during the early iterations. Thus, the imposed constraints substantially reduce the measurement overhead without compromising the ability of the adaptive framework to construct expressive, low-depth ans\"{a}tze. Here, the growth of $\mathcal{O}_{\rm pool}$ is governed primarily by the qubit connectivity and circuit design constraints, whereas its size is restrained by non-redundancy constraint. Although not shown here, we observed that adopting four or five operators per iteration does not alter the qualitative behavior presented here.

\section{Summary and Conclusion}
\label{Sec:Summary}

\par In this work, we emphasized the importance of tuning algorithmic frameworks according to circuit design constraints. In this context, we formulated adaptive variational algorithms in which constraints imposed by both the hardware platform and the circuit design, i.e. Hadamard test architecture, are incorporated into the gate pool. We showed that circuit design constraints substantially prune the hardware-aware gate pool and regulate its growth over successive iterations of adaptive algorithms. Consequently, this leads to a substantial reduction in the measurement overhead typically associated with adaptive algorithms. 

\par To assess how effectively the adaptive framework utilizes a constrained gate pool to construct expressive ans\"{a}tze, we investigated the ground state problem of the nonlinear Schr\"{o}dinger equation. In particular, we investigated two gate selection criteria: one based on the minimum cost value and other based on the maximum gradient value. For these problem instances, we demonstrated that adaptive algorithms effectively utilize the constrained gate pool to construct low-depth ans\"{a}tze without compromising expressivity. In general, our results highlight that incorporating circuit design constraints into the algorithmic framework provides a systematic route to a resource-friendly implementation of quantum algorithms.

\acknowledgments
\par This work is supported by the National Research Foundation, Singapore, and A*STAR under its Centre for Quantum Technologies Bridging Grant, and the EU HORIZON --- Project 101080085 --- QCFD. We thank Dr. Spyros Tserkis for a careful reading of the manuscript and helpful suggestions.

\section*{Data Availability Statement}
The data and codes cannot be made publicly available upon publication due to legal restrictions preventing unrestricted public distribution. The data and codes that support the findings of this study are available upon reasonable request from the authors.

\bibliography{adapt_NLSE_bib}

\end{document}